\documentclass[preprint2]{aastex}

\shorttitle{Stellar--mass--dependent disk structure}
\shortauthors{Sz\H{u}cs et al.}

\begin{document}

\title{Stellar--mass--dependent disk structure in coeval planet--forming disks}

\author{L\'{a}szl\'{o} Sz\H{u}cs}
\affil{Department of Experimental Physics, University of Szeged, Szeged, D\'{o}m t\'{e}r 9, 6720 Hungary}
\email{szucs@titan.physx.u-szeged.hu}

\author{D\'{a}niel Apai and Ilaria Pascucci\altaffilmark{1}}
\affil{Space Telescope Science Institute, 3700 San Martin Drive, Baltimore, MD 21218}
\altaffiltext{1}{Johns Hopkins University, Department of Physics \& Astronomy, 3400 N. Charles Street, Baltimore, MD 21218}
\email{apai@stsci.edu}
\email{pascucci@stsci.edu}

\and

\author{Cornelis P. Dullemond}
\affil{Max-Planck-Institut f\"ur Astronomie, Koenigstuhl 17, 69117 Heidelberg}
\email{dullemon@mpia.de}

\begin{abstract}
Previous studies suggest that the planet--forming disks around very--low--mass stars/brown dwarfs may be flatter than those around more massive stars, in contrast to model predictions of larger scale heights for gas--disks around lower--mass stars.
We conducted a statistically robust study to determine whether there is evidence for stellar--mass--dependent disk structure in planet--forming disks.
We find a statistically significant difference in the Spitzer/IRAC color distributions of disks around very--low--mass and low--mass stars all belonging to the same star-forming region, the Chamaeleon I star--forming region. We show that self consistently calculated disk models cannot fit the median spectral energy distributions (SEDs) of the two groups. These SEDs can
be only explained by flatter disk models, consistent with the effect of dust settling in disks. 
We find that relative to the disk structure predicted for flared disks the required reduction in disk scale height is anti--correlated with the stellar mass, i.e. disks around lower--mass stars are flatter. Our results show that the initial and boundary conditions of planet formation are stellar--mass--dependent, an important finding that must be considered in planet formation models.

\end{abstract}


\section{Introduction}
Circumstellar disks around young stars provide the raw material for planet formation. In addition, the evolution of protoplanetary disks sets the boundary conditions for the agglomeration of dust grains into planetesimals and for the assembly of planets from planetesimals
\citep[e.g. see reviews in][]{ApaiandLauretta2010}. Cool stars, with spectral types M0 or later, account 
for more than 80\% of the galactic stellar population. Their abundance --- combined with the suite of 
recent discoveries of planets around M--dwarfs \citep{Gaudi2008} --- suggest that cool stars are the typical planet hosts in the Galaxy. 
But how do these planetary systems differ from those around sun--like stars? 

Lacking observational data, some models of planet formation  assumed that disk properties are independent
of the stellar properties \citep[e.g.][]{Kornet2006}. Disk mass measurements, however, revealed that the disk/stellar
mass ratio is similar across a broad range of stellar masses \citep[e.g.][]{Klein2003,Andrews2005,Scholz2006,Bouy2008}.
This led to refined models that explained the observed low occurrence rate of gas giant planets
around M--stars \citep[][]{Laughlin2004,IdaLin2005,Johnson2007,Benz2008, Kennedy2008}. 
However, all other disk parameters are currently assumed to be essentially stellar--mass--independent. 

Recent observational studies, however, revealed stellar--mass--dependence for most disk properties. The typical lifetime of 
optically thick dust disks around cool stars and brown dwarfs seems to 
be at least a factor of two or three longer \citep[][]{Sterzik2004,Carpenter2006,Riaz2006} than around sun--like stars
\citep[see][ for a review]{PascucciTachibana2010}. Evidence mounts that stellar mass accretion rates scale as $\dot{M} \propto M^2$, resulting
in very--low--mass accretion rates at the low--end of the substellar mass spectrum \citep[e.g.][]{Natta2004,Mohanty2005,Muzerolle2005,Herczeg2009}.
Surprisingly, dust processing in the mid--infrared--emitting regions of brown dwarfs and low--mass stars is more advanced than in disks around sun--like stars of similar age \citep[e.g.][]{Apai2002,Apai2004,Apai2005,Kessler-Silacci2006,Bouy2008,Pascucci2009,Riaz2009}. Excitingly, Spitzer gas--line spectroscopy
revealed a prominent difference in the organic chemistry in disks around very cool stars and sun--like stars \citep[][]{Pascucci2009}.
In addition, several earlier studies have suggested that disks around cool stars and brown dwarfs may be flatter
than those around more massive stars \citep[e.g.][]{Apai2002,Pascucci2003,Apai2005}, although disk models predict
the opposite \citep[][]{Walker2004}.

The goal of this work is to provide a statistically robust evaluation of the hypothesis that the
lowest mass stars have, on average, flatter disks than their higher mass counterparts.
First, we present results from a statistical study of Spitzer/IRAC colors of the complete census of the 
Chamaeleon~I star-forming region. This comparison reveals a prominent difference between the
color distribution of the very--low-- and low--mass stars in the sample.
Next, we use a grid of simple disk models to demonstrate that the difference in the distributions of IRAC colors is well explained by the lower occurrence rate of flared disks around the lowest--mass stars. These findings further reinforce the stellar--mass--dependent evolution of protoplanetary disks, a key --- and yet unaccounted for --- factor to consider in planet formation models.

\begin{figure*}
\epsscale{2}
\plotone{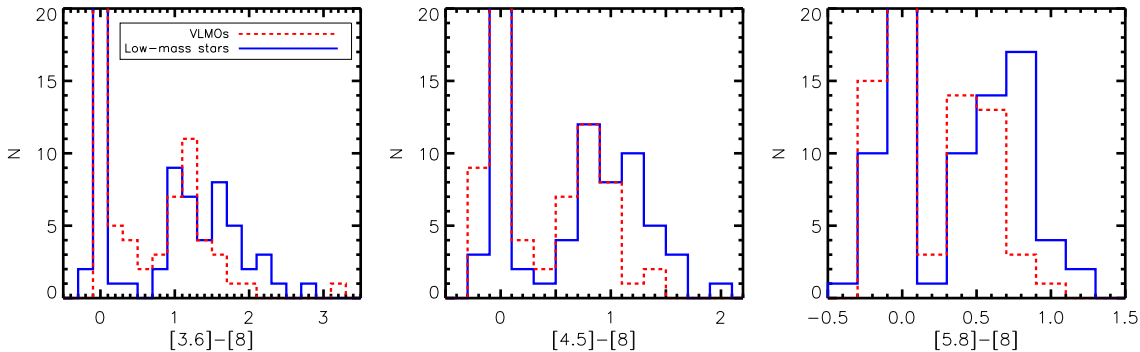}
\caption{The distribution of the IRAC--IRAC colors. The solid lines are the distributions of the low--mass stars, while the dashed lines represent the VLMOs. The peak close to 0 mag represents the stars without infrared excess. The distributions of the disk-bearing VLMOs are consistently shifted towards bluer colors on each diagram.\label{fig1}}
\end{figure*}

\section{Data Analysis}
In this study we focused on the Chamaeleon~I star--forming region (Cha~I, $\sim$200 known members \citealt{Luhman2008a,Luhman2008b}),
which provides an ideal set of targets. The membership of the cluster can be considered virtually complete for masses greater than $\sim$16 M$_{\rm J}$, providing an unbiased sample of roughly coeval, well--characterized stars  and disks \citep{Luhman2007,Apai2005,Pascucci2008,Pascucci2009,Natta2004,Riaz2009}. 
While the mean age of the very--low-- and low--mass stars may be slightly
different in Cha~I, the difference is thought to be lower than the estimated age spreads for each group \citep{Luhman2008c} and therefore can be assumed coeval for the purposes of this study.

We used photometric data from the Spitzer Space Telescope \citep{Werner2005}  obtained by the Infrared Array Camera \citep[IRAC,][]{Fazio2004} at 3.6, 4.5, 5.8, and 8 $\mu$m wavelengths and the Multiband Imaging Photometer for Spitzer \citep[MIPS,][]{Rieke2004} at 24 $\mu$m. 
The IRAC and MIPS observations were performed within a radius of $3^{\circ}$ from $\alpha=11^{h}07^{m}00^{s}$, $\delta=-77^{\circ}10^{'}00^{"}$. We complemented these data with $JHK_{\rm s}$ measurements from the Two Micron All Sky Survey \citep[2MASS,][]{Skrutskie2006}. 
We use the photometric data from \citet{Luhman2008a,Luhman2008b} and 2MASS \citep{Skrutskie2006} with the physical parameters of the members determined by \citet{Luhman2007,Luhman2008b}. For each source we adopted spectroscopically
determined  J--band extinction from \citet{Luhman2008a,Luhman2008b} and dereddened the measured magnitudes using the extinction law derived by the linear interpolation of $A(\lambda)/A(J)$ from \citet{Mathis1990}. For the purpose of comparing spectral energy distributions we calculated the flux densities in each band using zero points of the IRAC and MIPS instruments \citep{Cohen2003,Rieke2004,Reach2005}.
The spectral types of the stars and brown dwarfs in our sample range from B6.5 to M9.75 \citep{Luhman2007}. We divided these objects into two groups: A) low--mass stars (from G5 to M4.5, 110 stars); and, B) very--low--mass objects (VLMOs, from M4.75 to M9.5, 90 stars). The separation between the two groups corresponds to $T_{\rm eff}=3,170 K$.

\section{Results} \label{colordist}

We computed the IRAC colors for all objects as the difference between a magnitude at a short and at a longer wavelength.
Fig.~\ref{fig1} shows the IRAC--IRAC color distributions of the low--mass stars and VLMOs. Objects with colors close to zero are consistent
with pure photospheric emission, while objects with red colors (higher flux at longer wavelengths) have excess emission arising from a dust disk. \citet[][]{Hartmann2005} have evaluated IRAC--IRAC colors as disk indicators and have shown that the $[5.8]-[8] > 0.35$ criterion is efficient in discriminating between stars with and without disks. 
We adopt this criterion for our sample of Cha~I objects leading to 78 disk--bearing and 90 diskless stars. There are 32 stars that do not have either 5.8 or 8 $\mu$m measurements, and thus cannot be classified.

We find that disks in the VLMO group have bluer IRAC--IRAC colors than disks in the low--mass
sample (see Fig.~\ref{fig1}). To verify the statistical significance of this difference we performed a Kolmogorov--Smirnov test \citep{Vetterling2002} on the 47 low---mass and 31 VLMO disk--bearing members of the cluster.

In the case of the IRAC--IRAC color distributions the probability that the two samples are drawn from the same parent population are below 0.01 (see Table \ref{table1}). The low probabilities in all colors provide a solid proof for a
different color distributions in the two groups.
Furthermore, we also performed a K--S test for the IRAC--MIPS color distributions. For the IRAC--MIPS distributions the probability of drawing the two distributions from the same parent population is between 0.1114 and 0.0174, suggesting a possible difference, but not reaching the satisfactory confidence level.

\begin{deluxetable}{cccc}
\tablecolumns{4}
\tablewidth{0pt}
\tablenum{1}
\tabletypesize{\small}
\tablecaption{Probabilities that the IRAC colors of the disk bearing VLMOs and low--mass stars are drawn from the same parent population, as determined with a  Kolmogorov--Smirnov test.\label{table1}}
\tablehead{\colhead{IRAC Colors} &\colhead{[3.6]-[8]} & \colhead{[4.5]-[8]} & \colhead{[5.8]-[8]}}
\startdata
Probability & 0.00219 & 0.00023 & 0.00074 \\
\enddata
\end{deluxetable}

When interpreting this result, however, an important fact has to be kept in mind. While the IRAC maps are
sensitive to the photospheres of {\em all} stars and brown dwarfs in Cha~I, the MIPS measurements are sensitivity--limited
and are therefore biased. The detection limit of the MIPS measurements introduces a bias for the fainter VLMOs:  the fainter, less flared disks and/or diskless objects are often missing from the MIPS maps. The absence of the stars with less flared disks skews the IRAC--MIPS color distribution toward flared disks and reduces the observable difference in the IRAC--MIPS distributions of the samples. In spite of the biased MIPS photometry the K--S test still suggests a difference between the two samples.

We also find a difference in the IRAC--IRAC color distributions of the diskless stars. The color distributions of the diskless VLMOs are shifted towards redder colors, most likely due to strong molecular absorption bands in the atmospheres of the very--low--mass stars. Note, however, that this shift is to the {\em opposite direction} to the one seen in the disk--bearing sample.

There are three stars in our sample with IRAC colors consistent with pure photospheres ($\sim$0.0 mag), but with large [8]-[24] color ($>$2 mag). These sources could be stars with transition disks \citep[e.g.][for the definition]{Storm1989}. These stars do not alter our statistics significantly, but they are interesting for further studies. The 2MASS identifiers of these stars are J11022491-7733357, J11071330-7743498, and J11124268-7722230.

In summary, the K--S test convincingly demonstrates that the color distribution of the disks around the VLMOs and low--mass stars is different: VLMOs have disks with distinctly bluer IRAC colors. 
Earlier studies of small disk samples have hinted on such differences, often thought to be linked to the effects of dust settling on disks \citep{Apai2004,Apai2005,Kessler-Silacci2006,Dullemond2004b,Furlan2008}. 
Seeking an explanation for the observed color difference, in the next section we will use a grid of 
simple disk models to verify this explanation and explore whether other possibilities are also viable.

\begin{figure}[!ht]
\epsscale{1}
\plotone{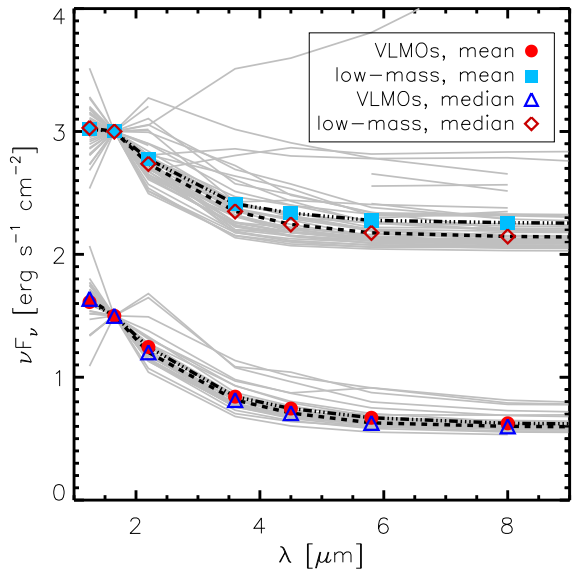}
\caption{Spectral Energy Distributions of the disk bearing members of the VLMO and low--mass samples normalized to the H band.\label{fig2}}
\end{figure}

\section{Disk Model Grid} \label{modeling}
Fig. \ref{fig2} shows the spectral energy distributions (SEDs) of all disk--bearing members of the low--mass and VLMO groups, normalized to their H--band flux density. 
The 24 $\mu$m flux densities are not shown on this figure, because of the biased subset of the VLMO groups, as discussed in Section 3.

While the {\em median} SEDs of the two groups have similar shape, the {\em mean} SEDs are different. In the case of the VLMOs the medians and means of the fluxes have 
similar values at a given wavelength, while in the case of the low--mass sample the mean IRAC values are higher than the median IRAC values. This indicates 
more significant broadening of the flux density distributions towards higher fluxes than in the case of the VLMOs (see Fig.~\ref{fig2}). The higher broadening of flux density distributions of the 
low--mass sample is consistent with the results of the K--S tests: the flared disk to flatter disk ratio is higher in the case of the low--mass stars than for VLMOs.

For modeling the disks we will use the median SED of the unbiased low--mass sample (including the 24 $\mu$m wavelength) and the median SED of the unbiased VLMO sample (JHK and IRAC bands), with the median of the biased 24 $\mu$m flux density as an {\em upper limit}. The median 24~$\mu$m flux density of the sensitivity--limited (and therefore biased) VLMO sample provides an {\em upper limit} for the 24~$\mu$m median flux of the VLMOs. If the observations
were not missing the fainter sources, the  median 24~$\mu$m flux density would be lower.

In order to quantitatively interpret the differences observed in the IRAC colors and in the distribution of the SEDs we used a grid
of disk models with and without dust settling. We applied the RADMC radiative transfer code \citep{Dullemond2004a}, 
which can determine the vertical density structure of the disk self--consistently, assuming perfect dust--gas coupling and hydrostatic equilibrium for the gas.
The RADMC is complemented with RAYTRACE, a post-processing tool to calculate the emerging SED. 
The primary input parameters of the RADMC/RAYTRACE codes are the temperature ($T_{\rm eff}$), luminosity ($L_{\rm bol}$), the mass ($M_{\rm star}$) of the central star, and the inner radius ($R_{\rm in}$), outer radius ($R_{\rm disk}$), inclination ($i$), and mass ($M_{\rm disk}$) of the disk.
Another key parameter of the RADMC code is the number of photons used in the radiative transfer. 
The low photon number could cause fluctuations of the mid-plane temperature of the disk, especially at the inner 0.5 AU where the near and mid--infrared radiation of the dust is coming from. The fluctuation could be reduced by a photon diffusion algorithm \citep{Min2009}, but it can not be used in every case. Thus, the photon number should be set carefully.
The changing of the pressure scale height ($H_{p}$) with the radius  describes the disk flaring. The pressure scale height in RADMC is expressed in the unit of the distance from the star ($H/a$, where $a$ is the radial distance from the star) and can be set externally. A flatter disk has a smaller $H/a$ than a flared disk at the same emitting region.

We adopted the median effective temperature and bolometric luminosity of the stars of the two groups, as determined in \citet{Luhman2007,Luhman2008b}. 
The stellar masses were derived from the theoretical evolutionary models of \citet{Baraffe1998}. The adopted stellar parameters 
of the VLMOs and the low--mass sample are shown in Table~\ref{table2}.

\begin{deluxetable}{lrr}
\tablewidth{0pt}
\tablenum{2}
\tabletypesize{\small}
\tablecaption{Adopted stellar parameters.\label{table2}}
\tablehead{\colhead{Parameters} & \colhead{VLMO} & \colhead{Low-mass}}
\startdata
$T_{\rm eff} $[K] & 3,024 & 3,669 \\
$L_{\rm bol} $ [$L_{\rm \odot}$] & 0.03 & 0.39 \\
$M_{\rm star} $ [$M_{\rm \odot}$] & 0.08 & 0.60 \\
\enddata
\end{deluxetable}

First, we calculated flared disk models by setting the vertical structure self--consistently, i.e. assuming perfect dust--gas coupling and hydrostatic equilibrium for the gas. The model grid parameters and their range are as follows: $R_{\rm in}$ from 2 $R_{\rm star}$ to 7 $R_{\rm star}$, $M_{\rm disk}$ from 5$\times10^{-3}M_{\rm star}$ to 0.05 $M_{\rm star}$. We kept the $R_{\rm disk}$ constant: in the case of the VLMO sample it was 50 AU, while in the case of the low--mass sample it was 100 AU. The pressure scale height was calculated by the code in 10 iteration steps. The number of photons was set to 300,000; this photon budget was found to be sufficient to achieve a smooth mid--plane temperature profile even in the inner regions of the disks, when using the diffusion algorithm.
We calculated SEDs for inclinations ranging from $40^{\circ}$ to $70^{\circ}$ with $5^{\circ}$ steps. 

\begin{figure*}
\epsscale{1.5}
\plotone{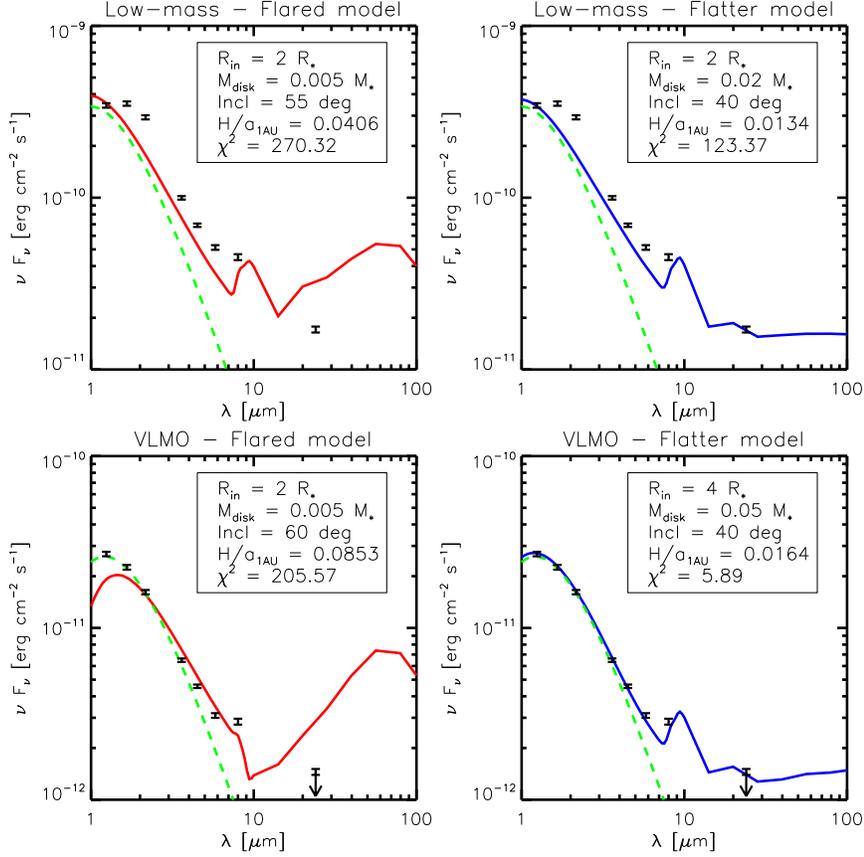}
\caption{Best--fit models with dust settling explain better the median SEDs than the best--fit fully flared models. The upper panels shows the best--fitting fully flared (upper left) and flatter (upper right) model of the low-mass group, the lower panels shows the best--fitting fully flared (lower left) and flatter (lower right) model of the VLMOs. Dashed lines represents the black body SEDs calculated with the adopted stellar parameters. The solid lines shows the best--fitting models. Note, that the relatively high $\chi^2$ values for certain models are partly due to the non--black body spectrum of the stellar photosphere and the fine structure of the silicate emission feature, none of which is included in our models.\label{fig3}}
\end{figure*}

We performed a reduced $\chi^{2}$ fitting for the median SEDs to identify the best--fitting model. The errors were calculated by $F_{\rm err, med.}(\lambda)=median(F_{\rm err}(\lambda)/F(\lambda))\times median(F(\lambda))$. The best--fitting model for the low-mass group is $R_{\rm in}=2R_{\rm star}$, $M_{\rm disk}=5\times10^{-3}M_{\rm star}$, and $i=55^{\circ}$ with $\chi^2$ = 270.32. In the case of the VLMOs the best--fit is $R_{\rm in}=2R_{\rm star}$, $M_{\rm disk}=5\times10^{-3}$M$_{\rm star}$, and $i=60^{\circ}$ $\chi^2$ = 205.57. The $H/a$ pressure scale height of a fully flared, radiative equilibrium disk is given by $H/a = \sqrt[]{{k T a^3} / {\mu_{\rm gas} m_{\rm p} G M_{\rm star}}} / a$, where $T$ is the mid-plane temperature at $a$ radius of the disk, $\mu_{\rm gas}$ is the mean molecular weight of gas and $m_{\rm p}$ is the photon mass \citep[see eq. 6 in][]{Dullemond2004b}. Thus, the $H/a$ at $a = 1$ AU for the best--fitting models are 0.041 (low--mass) and 0.085 (VLMOs). The best--fitting fully flared models are shown on Fig.~\ref{fig3} (upper and lower left panels).

The fully flared models could not fit the median SEDs of the two groups satisfactorily within the parameter ranges of our model grid. The best--fitting model of the low--mass group overpredicts the median observed 24 $\mu$m fluxes, suggesting that a flatter disk is required. On the other hand the model underpredicts the SED at $H, K_{\rm s}$ and IRAC wavelengths, this effect likely resulting from the absence of accretion in our models. To verify this we compared the amount of $H$ and $K_{\rm s}$ band flux excess to the $H$ and $K_{\rm s}$ fluxes of the accretion disk models of \citet{D'Alessio2004}. The amount of excess at $H$ and $K_{\rm s}$ bands could be reproduced by models with an
accretion rate of $\sim10^{-8} M_{\rm sol}/yr$, a typical mass accretion rate for a few Myr--old star with disk. Thus, the poor fit of the shorter infrared wavelengths is likely caused by the lack of accretion in our models. 

Fully flared disks provide an even worse fit for the VLMO group. Fitting the $K_{\rm s}$ and IRAC data points requires unrealistically low disk mass and very extreme inclinations; on the other hand, 
these extreme models predict more flux at 24 $\mu$m than the observed upper limit.

In order to verify whether dust settling and flatter disks may explain the observed disk structures, we expanded our model grid with an additional parameter. We express dust settling by allowing the pressure scale height at $R_{\rm disk}$ to vary between 0.01 and 0.3. The flatter disks require more model photons: we set the number of photons to 2.5 million to ensure continuous  and smooth mid--plane temperature profile. On this expanded grid the best--fit model for the low--mass sample is $R_{\rm in} = 2 R_{\rm star}$, $M_{\rm disk} = 0.02 M_{\rm star}$, $i = 40^{\circ}$, and $H/a = 0.05$ with $\chi^2$ = 123.37. The best--fit model for the VLMOs is $R_{\rm in} = 4 R_{\rm star}$, $M_{\rm disk}= 0.05 M_{\rm star}$, $i=40^{\circ}$, and $H/a=0.05$ with $\chi^2$ = 5.89. Thus, allowing dust settling and flatter models provides much better fits and disk masses more consistent with those observed \citep[see Fig. 8 in][]{Andrews2005} and inclinations, close to what is expected from a large sample of randomly oriented disks. The best--fit models from this grid are shown in Fig.~\ref{fig3} (upper and lower right panels).

Our expanded grid reveals that flatter disks are a much better match to the SEDs than flared ones. In the case of the low--mass group, the best--fit flatter disk model (Fig.~\ref{fig3} upper right panel) underestimates the $H, K_{\rm s}$, and IRAC fluxes similarly to the fully flared model, but it fits the 24~$\mu$m flux well. In the case of the VLMOs the best--fit model fits the observations almost perfectly. There is a little deviation at the IRAC [8] band. This non--perfect fit is expected, because the flux density at wavelengths close to 9 $\mu$m are strongly affected by the shape of the silicate emission feature, a parameter not varied in our models. The 24~$\mu$m flux of the models is a sensitive proxy of the pressure scale height (and flaring) of the disk: a lower scale height results in a lower 24~$\mu$m flux. Thus, the upper limit of the 24~$\mu$m flux of the VLMOs translates to an {\em upper limit} on the pressure scale height.

To compare the pressure scale heights of the best--fit flatter models, we calculated the scale heights of the model disks at the radii where the observed IRAC and MIPS fluxes emerge. As a first approximation the protoplanetary disks could be described as a black body radiator with its temperature decreasing with the radius. According to Wien's displacement law we calculated the temperatures of the black bodies which have the maximum of radiation at the IRAC and MIPS wavelengths and interpolated the mid-plane temperature profile of the disk to determine the radius where the radiation emerges. Following \citet{Chiang1997} we assumed a flaring angle of 2/7 and calculated the pressure scale height at these radii ($a$) by $H/a = [H/a]_{\rm 0} \times (\frac{a}{a_{\rm 0}})^{2/7}$, where $a_{\rm 0}$ and $[H/a]_{\rm 0}$ are the outer radius and the pressure scale height at the outer radius. 

Table \ref{table3} summarizes the pressure scale heights of the best--fitting fully flared low--mass, VLMO and flatter low--mass, VLMO disk models at the
regions of the disks where the 3.6, 4.5, 5.8, 8, and 24 $\mu$m radiation emerge. The scale height of the VLMO disks is reduced by about a factor of 4 relative to their predicted structure. In contrast, that for the low--mass group is reduced by only a factor of ~2.

\begin{deluxetable}{lrrrrr}
\tablewidth{0pt}
\tablenum{3}
\tabletypesize{\small}
\tablecaption{Pressure scale heights of the best--fit disks at regions that dominate emission in various wavelengths\label{table3}}
\tablehead{\colhead{Group} & \colhead{3.6 $\mu$m} & \colhead{4.5 $\mu$m} & \colhead{5.8 $\mu$m} & \colhead{8 $\mu$m} & \colhead{24 $\mu$m}}
\newcommand\T{\rule{0pt}{3.1ex}}
\newcommand\B{\rule[-1.7ex]{0pt}{0pt}} 
\startdata

\multicolumn{6}{c}{Fully flared models} \B \\
Low--mass & 0.0112 & 0.0120 & 0.0128 & 0.0135 & 0.0178 \\
VLMO & 0.0181 & 0.0188 & 0.0197 & 0.0215  & 0.0281\B \\ \hline

\multicolumn{6}{c}{Flatter models} \T \B \\
Low--mass & 0.0046 & 0.0050 & 0.0054 & 0.0062 & 0.0088 \\
VLMO & $<$0.0048 & $<$0.0050 & $<$0.0053 & $<$0.0057 & $<$0.0082\B \\ \hline

\multicolumn{6}{c}{Relative pressure scale heights} \T \B \\
Low--mass & 2.42 & 2.43 & 2.37 & 2.18 & 2.02 \\
VLMO & $>$3.79 & $>$3.75 & $>$3.73 & $>$3.76 & $>$3.42 \\

\enddata

\end{deluxetable}

Our SED modeling also illustrates the power of mid-- to far--infrared photometry in accurately determining the disk geometry and
following the settling of dust and its impact on disk evolution. The recently launched Herschel Space Observatory provides a new
opportunity to study disks at these hard--to--access wavelengths. In the following we point out that Herschel is capable of mapping
the SED of typical VLMO disks in just minutes. We used our model SEDs to estimate the flux density at the 60-85 $\mu$m band of the Photodetector Array Camera and Spectrometer \citep[PACS,][]{Poglitsch2008} of the Herschel Space Observatory. The prediction of flux density at this band for the best--fitting fully flared low--mass model is 1278 mJy, while for the fully flared VLMO model it is 117 mJy. The best--fitting flatter low--mass model predicts 424 mJy, while the best--fitting VLMO flatter model predicts 36 mJy. The instrument is capable of 8 $\sigma$ detection in the faintest case with $\sim$11 minute exposure time. Thus, PACS measurements could distinguish between the fully flared and flatter geometries.

\section{Discussion}

Using colors derived from Spitzer photometry, median SEDs, and best--fit disk models we showed that the flaring
of disks around young low--mass and very--low--mass stars (VLMOs) is statistically different and that disks around cooler stars
are flatter. While previous studies by our group \citep[e.g.][]{Apai2005,Pascucci2009} and others already hinted at such a difference, its 
statistical significance has not yet been verified. Dust settling -- a physical process that leads to flatter disks -- enhances solid surface density in the disk
mid--plane and is an early, important step toward planet formation.  

Our study shows that in coeval disks this process is more efficient around lower--mass stars, an effect that has neither been predicted 
nor easily understood --- in fact, theoretical works predict the {\em opposite} due to the lower gravity of the central star \citep[see, e.g. ][]{Walker2004}. 

Although developing a detailed, quantitative model explaining the stellar--mass--dependent dust settling is beyond the scope
of the current paper, we venture to speculate about possible explanations
of the observations. Because the scale height of the fine dust is set by the dynamical coupling of dust and gas processes that lead to differences in the evolution of the gas disk, the dust disk or the dust--to--gas coupling may all lead to the observed stellar type--dependent disk structures.

A lower gas surface density in disks around cooler stars may provide a straightforward explanation for lower dust equilibrium scale heights, i.e. flatter disks. This model would require that either disks around cool stars lose their gas faster or that lower-mass stars form from cloud cores with a lower--than--average gas to dust ratio.

Other possible explanations include a weaker turbulence in disks around lower-mass stars, leading to weaker coupling
between the dust and gas components. As turbulence and viscosity also directly determines the average accretion rate through a disk, future studies may be able to find a consistent framework for connecting variations in disk structure and accretion with spectral type to systematic differences in disk viscosity. Because the origin of turbulence (and viscosity) in protoplanetary disks is poorly understood, observations such as those presented in this paper may provide indirect clues constraining on these long-standing questions.

However, more complex explanations are also possible. Detailed simulations of grain-grain collisions and their impact on the overall disk structure and spectral energy distributions have been discussed in the literature extensively, including the effects of collisions, fragmentation,
inward drift \citep[e.g.][]{Weidenschilling1997,Dominik2008,Birnstiel2009}. Supported by
laboratory studies of grain collision \citep[e.g.][]{Blum2008} the models find a very efficient and rapid grain growth
in disks on time scales shorter than the disk lifetime. Indeed, most  models would lead to
a rapid loss of the fine due component, in stark contrast to the observed large frequency of dust disks at even 1--3 Myr ages.
This suggests that processes other than simple grain--grain collisions and fragmentation are involved in defining
the grain size distribution in disks \citep[see also][]{PontoppidanBrearley2010}.  

We also point out that the flaring disk surface is defined by a small mass fraction of the dust in the disk; and disks around the
coolest stars and brown dwarfs will be about an order of magnitude less massive than those around
sun–like stars. However, the optical depths probed by infrared observations will be independent of the stellar spectral
type. In addition, if the dust component is in equilibrium then the scale height of the smaller dust component will be
larger than that of the larger grains. Thus, the observations at the same wavelengths may probe deeper in the less
massive disks around cool stars and lead to a larger deduced typical grain size and a flatter disk, even if the grain 
size distribution are identical in the two disks. 

The observed differences in disk structure is one of the first steps toward distinguishing between the interesting possibilities
identified above. Even more importantly, however, these results question the assertion that  
the initial conditions for planet formation are not stellar mass-dependent, frequently made in current planet formation
models. The observed difference in the disk structure is a very important clue that joins the ranks of differences observed in disk mass \citep[e.g.][]{Klein2003,Scholz2006,Williams2007}, dust properties \citep[][]{Apai2005,Pascucci2009}, accretion rates \citep[][]{Natta2004,Muzerolle2005,Mohanty2005}. Most recently, statistically significant stellar--mass--dependent differences have been identified even in the organic chemistry of disks \citep[][]{Pascucci2009}.

The evidence presented here on stellar--mass--dependent disk structures completes the picture and further emphasizes the need for disk models that naturally reproduce these stellar--mass--dependent disk properties and disk evolution. Only with such models we will be able to develop predictive, quantitative models for planet formation and understand the diversity of planetary systems.

\section{Summary}

Using Spitzer/IRAC colors of a complete sample of coeval young stars in the Cha~I star--forming regions we found that
the IRAC colors of disks around the lower--mass and higher--mass stars are statistically different: disks around cooler
stars are bluer, i.e. their SEDs have a less steep slope beyond 4~$\mu$m. The same difference is seen in the median SEDs
of the two samples. Using a grid of radiative transfer disk models we found that: A) fully flared disks are not consistent with the median
SEDs; B) disks in the lower--mass sample require a much stronger reduction of the disk scale height relative to the fully flared disk geometry than those in
the higher--mass sample. In short, the Spitzer IRAC data reveals
a stellar--mass--dependent disk structure and identifies more efficient dust settling in disks around lower--mass stars than in
their higher--mass counterparts. These findings demonstrate that the initial and boundary conditions of planet formation around
cool stars are different from those around sun--like stars.

\acknowledgments
L.Sz. acknowledges the Space Astronomy Summer Program of the Space Telescope Science Institute and support through the
Spitzer Data Analysis grant 1348621.
This work has also been supported by the Hungarian OTKA grant K76816.
We are grateful to Kevin Luhman for providing us the IRAC/MIPS photometric data and stellar properties of the members of Cha I. We thank Paul Harvey for the the helpful review that helped to improve the manuscript.

\end{document}